\documentclass[prd,twocolumn,amsmath,amssymb,floatfix,superscriptaddress,nofootinbib,preprintnumbers]{revtex4-1}

\usepackage{amsmath}
\usepackage{amssymb}
\usepackage{graphicx}
\usepackage{color}
\usepackage{hyperref}
\usepackage{subcaption}
\usepackage{epstopdf}
\usepackage{mathrsfs}  
\usepackage[utf8]{inputenc}
\usepackage{natbib}
\usepackage{multirow}

\newcommand{\dd}{\mathrm{d}}
\newcommand{\vv}{\vec}
\renewcommand{\[}{\begin{equation}}
\renewcommand{\]}{\end{equation}}

\def\T{\Theta_{\gamma b}}
\def\tb{\tau_{\chi}}
\def\tc{\tau_c}

\begin{document}

\title{Probing sub-GeV Dark Matter-Baryon Scattering with Cosmological Observables}

\author{Weishuang Linda Xu, Cora Dvorkin, Andrew Chael}

\affiliation{Harvard University, Department of Physics, \\Cambridge, MA 02138, USA}

\begin{abstract}
We derive new limits on the elastic scattering cross-section between baryons and dark matter using Cosmic Microwave Background data from the Planck satellite and measurements of the Lyman-alpha forest flux power spectrum from the Sloan Digital Sky Survey.  Our analysis addresses generic cross sections of the form $\sigma \propto v^n$, where $v$ is the dark matter-baryon relative velocity, allowing for constraints on the cross section independent of specific particle physics models.   We include high-$\ell$ polarization data from Planck in our analysis, improving over previous constraints.  We apply a more careful treatment of dark matter thermal evolution than previously done, allowing us to extend our constraints down to dark matter masses of $\sim$MeV.   We show in this work that cosmological probes are complementary to current direct detection and astrophysical searches.
\end{abstract}

\maketitle

\section{Introduction}

The standard paradigm for dark matter (DM) in contemporary cosmology is that it is cold and collisionless, interacting only gravitationally with Standard Model particles. While successful on large scales \cite{Ade:2015xua}, the data still allow for a rich variety of non-minimal models \cite{Preskill:1982cy, Ali-Haimoud:2016mbv, Jungman:1995df,  Feng:2009mn, Kaplan:2009ag, Dolgov:2002wy, Fan:2013yva}, and the particle nature of dark matter is still very much unknown. In particular, tensions between observations and cold dark matter (CDM)-based simulations on galaxy scales \cite{Weinberg:2013aya, Bullock:2017xww} provide motivation to explore new types of DM interactions that are not accessed by direct searches: the  ``core-cusp'' \cite{Spergel:1999mh,Walker:2011zu, Salucci:2012,Donato:2009} , ``missing satellite'' \cite{Klypin:1999uc, Moore:1999nt}, and ``too big to fail'' \cite{BoylanKolchin:2011de, Papastergis:2014aba} problems at the small-scale indicate that dwarf galaxies are fewer and less centrally dense than predicted by $\Lambda$CDM simulations. While these problems may not necessarily require new physics \cite{Brooks:2012vi, Read:2017lvq, Drlica-Wagner:2015ufc, Arraki:2012bu}, they nevertheless provide motivation to look at cosmologies beyond the CDM scenario. 

In this work, we explore the cosmological effects of dark matter interacting with baryons via elastic scattering. We specifically investigate scenarios in which the DM-proton elastic scattering cross-section $\sigma$ scales effectively as a power-law of the baryon-dark matter relative velocity $\sigma = \sigma_0 v^n$, and we provide constraints independent of the underlying particle model.  This type of relation naturally occurs in a number of different models, and we will focus our analysis on several values of $n$ that are particularly well-motivated: $n = \{ -4, -2, -1, 0 ,2 \}$, which can for instance correspond to DM with fractional electric charge ($n=-4$) \cite{Melchiorri:2007sq}, a Yukawa potential (a massive-boson exchange) ($n=-1$) \cite{ArkaniHamed:2008qn, Buckley:2009in} , velocity-independent scattering ($n=0$) \cite{Chen:2002yh} , and dark matter with electric and magnetic dipole moments ($n=\pm2$) \cite{Sigurdson:2004zp}.

Thermal coupling between DM and baryons in early times dampens the growth of fluctuations in the DM fluid and modifies the baryon relative velocity. The resulting power suppression on small scales and acoustic peak shift in the Cosmic Microwave Background (CMB) temperature and polarization power spectra, as well as the suppression of the matter power spectrum, allow us to constrain this type of interaction.
We use measurements of the CMB temperature and polarization power spectra by the {\it Planck} satellite (2015 results), and the Lyman-$\alpha$ forest flux power spectrum measurements by the {\it Sloan Digital Sky Survey} (SDSS) to obtain limits on DM-baryon elastic scattering.  Similar constraints have been considered also in Refs.\cite{Boehm:2004th, Cyburt:2002uw, Dvorkin:2013cea}; specifically velocity-independent scattering has been investigated in Refs. \cite{Chen:2002yh, Boehm:2001hm,Gluscevic:2017ywp} and millicharged DM in Refs. \cite{Dolgov:2013una,  Dubovsky:2003yn, Davidson:2000hf, Cline:2012is, McDermott:2010pa, Vogel:2013raa,Agrawal:2017pnb,Kamada:2016qjo}. Additional constraints on DM interactions have been derived from spectral distortions \cite{Ali-Haimoud:2015pwa, Diacoumis:2017hff}, galaxy clusters \cite{Chuzhoy:2004bc, Hu:2007ai, Qin:2001hh}, gravitational lensing \cite{Natarajan:2002cw, Markevitch:2003at}, the thermal history of the intergalactic medium \cite{Munoz:2017qpy, Cirelli:2009bb}, 21 cm observations \cite{Tashiro:2014}, indirect detection and gamma-rays  \cite{TheFermi-LAT:2017vmf, Mack:2012ju, Daylan:2014rsa, Goodenough:2009gk, Elor:2015bho, Madhavacheril:2013cna}, and direct detection searches \cite{Xiao:2017vys, Akerib:2016vxi, Agnese:2017njq, Angle:2007uj, Aalseth:2012if, Amole:2017dex, Angloher:2015ewa, Angloher:2017sxg}.

We extend previous work done in Ref. \cite{Dvorkin:2013cea} by applying our analysis to lower-mass dark matter particles, down to order $\sim$ MeV, restricting specifically to non-relativistic interactions with protons, and by including high-$\ell$ CMB polarization data from the Planck 2015 release. MeV-scale dark matter has previously been considered in Refs. \cite{Choudhury:2017osc, Hooper:2007tu,  Borodatchenkova:2005ct, Boehm:2003bt, Fichet:2017bng, Bertuzzo:2017lwt}. Our approach is particularly interesting for the $n=0$ scenario given its complementarity to current direct detection searches that generally target higher DM masses due to kinematic considerations. We will specifically compare to recent constraints from direct detection experiments \cite{Akerib:2016vxi, Angloher:2015ewa, Angloher:2017sxg, Erickcek:2007jv, Aprile:2017iyp} to illustrate this. For the $n=-4$ scenario, constraints on millicharged dark matter have been primarily derived from astrophysical sources and collider experiments \cite{Vinyoles:2015khy,Vogel:2013raa,Prinz:1998ua,Jaeckel:2012yz}. Our results are complementary to those.  

This paper is organized as follows: we review the modified Boltzmann equations including DM-baryon scattering in Section~\ref{BE} and the equations governing DM and baryon temperature evolution in Section~\ref{TE}.  A more detailed treatment, as well as the evolution equations under tight-coupling approximation, can be found in the Appendix. Our numerical results are presented in Section~\ref{NR}, and we discuss in detail the improvement for the $n=-4$ scenario from including CMB polarization anisotropy data in Section~\ref{PL}.  In Section~\ref{MS} we provide an extrapolation of our MCMC results applicable to all DM masses $\gtrsim$ 1 MeV.  In Section~\ref{DD} we compare our results for velocity- and spin-independent scattering to limits from direct detection experiments. Likewise, in Section~\ref{MC} we compare our results for millicharged DM to existing constraints from other sources. 

\section{Boltzmann Equations \label{BE}} 

We review the modifications to the dark matter and baryon Boltzmann equations to account for DM-baryon scattering presented in Ref. \cite{Dvorkin:2013cea}.  We work in a modified synchronous gauge, allowing for a nonzero peculiar velocity of dark matter $\vec V_\chi$ when scattering is turned on.  For a given Fourier mode $k$, the density fluctuations $\delta_\chi$ and $\delta_b$  and velocity divergences  $\theta_\chi$ and $\theta_b$ of the DM and baryon fluids obey the following equations

\begin{eqnarray} 
{\dot \delta_\chi} &=& -\theta_\chi - \frac{\dot h}{2}, \\ 
{\dot \delta_b}&=& -\theta_b- \frac{\dot h}{2},  \\
{\dot \theta_\chi} &=& -\frac{\dot a}{a}\theta_\chi  + c^2_\chi k^2 \delta_\chi + R_\chi (\theta_b - \theta_\chi ), \\
{\dot \theta_b} &=& -\frac{\dot a}{a}\theta_b + c^2_b k^2 \delta_b + R_\gamma (\theta_\gamma - \theta_b) \nonumber \\ \phantom{\frac{\dd \theta_b}{\dd \eta}} && \phantom{-\frac{\dot a}{a}\theta_b + c^2_b k^2 \delta_b} + \frac{\rho_\chi}{\rho_b} R_\chi (\theta_\chi- \theta_b),
\end{eqnarray} 
where overdots denote derivatives with respect to conformal time, $h$ denotes the metric perturbation, $c_\chi$ and $c_b$ refer respectively to the DM and baryon sound speeds,  $R_\gamma$ is the momentum-transfer rate for baryon-photon coupling (as set by Thompson scattering), and $R_\chi$ is that for DM-baryon coupling.

The momentum-exchange rate $R_\chi$ is set by the cross-section $\sigma_0$ and power-law index $n$ as 
\begin{equation} 
R_\chi =  \frac{a \rho_b \sigma_0 c_n }{m_\chi + m_b} \left(  \frac{T_b}{m_b} + \frac{T_\chi}{m_\chi} + \frac{V^2_{RMS}}{3}\right)^{\frac{n+1}{2}} \mathcal{F}_{He}, \label{rchi}
 \end{equation}
where $T_{b (\chi)}$ and $m_{b (\chi)}$ are the baryon (DM) temperature and particle masses and $c_n$ is an $n$-dependent constant tabulated in Table~\ref{cns} in the Appendix. 
This expression is valid to leading order for both early times ($z>10^4$), where the thermal velocity dispersion dominates over the DM bulk velocity, and at late times where the peculiar velocity dominates.  

Following Ref. \cite{Dvorkin:2013cea}, we write $V^2_{RMS}$, the averaged (with respect to the primordial curvature perturbation) value of  $V^2_\chi$ as
\begin{equation} 
V^2_{RMS} \equiv \langle V^2_\chi \rangle 
\simeq  \begin{cases} 10^{-8} & z>10^3 \\ 10^{-8} \left(\frac{(1+z)}{10^3}\right)^2  & z\leq 10^3 \end{cases} . \label{eq:vrms} 
\end{equation}
The factor $\mathcal{F}_{He}$ accounts for the significant fraction of helium in the baryon population and can encode different dynamics for scattering off Helium. For the case of no scattering between DM and Helium this is simply $\mathcal{F}_{He} = 1 - Y_{He} \approx 0.76$. 

A derivation of the form of $R_\chi$ from DM-baryon drag, and a detailed treatment of the Boltzmann equations in the tight coupling regime is given in the Appendix. 

\section{Thermal evolution of DM\label{TE}} 

The temperature evolution of the DM and baryon fluids with DM-proton scattering is given by

\begin{eqnarray}
\label{eq:tempev}
\dot T_\chi &=& -2 \frac{\dot a}{a} T_\chi    +   \frac{2 m_\chi}{m_\chi + m_b} R_\chi (T_b - T_\chi), \\
\dot T_b &=& -2 \frac{\dot a}{a} T_b    +   \frac{2 \mu_b}{m_e} R_\gamma (T_\gamma - T_b)  \nonumber \\  
\phantom{\dot T_b} &&  \phantom{-2 \frac{\dot a}{a} T_b } +  {\rho_\chi\over\rho_b}\frac{2 \mu_b}{m_\chi + m_b} R_\chi (T_\chi - T_b),
\label{eq:barytempev}
\end{eqnarray}
where, again, overdots denote derivative with respect to conformal time. Here $ \mu_b$ denotes the mean molecular weight for the baryons, $\mu_b=m_H(n_H+4n_{H_e})/(n_H+n_{H_e}+n_e)$. 

In Ref. \cite{Dvorkin:2013cea}, the authors assumed that the DM fluid remains thermally coupled with baryons until late times since for DM particle masses heavier than the mass of a proton the corrections due to a temperature difference between baryons and dark matter are suppressed. We relax this assumption to extend the validity of our results to lower DM masses.

Ref. \cite{Tashiro:2014} explored numerical solutions to Eq.~\ref{eq:tempev} for different values of $n$ and $\sigma$ and calculated the effect on the 21 cm power spectrum. Ref. \cite{Munoz:2015bca} extended this calculation to be valid at late times when  the peculiar velocity dominates over the DM-baryon thermal velocity.  Due to the baryon-dark matter interaction, the baryons are cooled relative to $\Lambda$CDM evolution after decoupling from photons. Numerical solutions to Eq.~\ref{eq:tempev} show that for $n>-4$, dark matter decouples from the photon-baryon fluid at high redshift for reasonable values of $\sigma$. Thus, for $n>-4$, rather than solving the full temperature evolution equations, we can apply the simple approximation that the dark matter remains thermally coupled with the baryon-photon fluid until the rate of expansion exceeds the rate of scattering, at which point the DM component suddenly decouples and evolves adiabatically:

\begin{equation}
T_\chi  = \begin{cases} T_b, & R_\chi \frac{m_\chi }{m_\chi + m_b} > aH \\   T_{dec}\left({a_{dec} \over a}\right)^2, & R_\chi \frac{m_\chi }{m_\chi + m_b} < aH,\end{cases}
\label{eq:decoupling}
\end{equation}
where the subscript ``{\it dec}" denotes the time at which dark matter decouples from the photons and baryons.

For $n=-4$, the dark matter-baryon coupling strength \emph{increases} with time, and numerical solutions to Eq.~\ref{eq:tempev} show that the dark matter instead recouples to baryons at late times for sufficiently strong scattering.  In this case, a sudden decoupling approximation is no longer valid.  In fact, since DM and baryons do not thermally couple via this interaction at early times at all, the initial condition of $T_\chi$ becomes model dependent.  Here, we assume a WIMP-like scenario: at higher energies DM annihilates to baryons through weak-scale interactions. After freeze-out, the DM temperature evolves adiabatically until the $n=-4$ scattering (e.g. as induced by millicharge) becomes important. The DM temperature initial condition is then set by 
\begin{equation}
T_\chi (z) = T_b(z) \qquad \text{at}\qquad H(z) = \rho_\chi/m_\chi \langle \sigma_w v \rangle ,
\end{equation}
where we take a weak scale cross-section $\langle \sigma_w v\rangle \sim 10^{-26}\text{ cm}^3/\text{s}$, and choose $\rho_\chi$ such that it matches the DM relic abundance today. However, in practice, the $n=-4$ scenario is sufficiently constrained that, at redshifts where the modes measured by the CMB and Lyman-$\alpha$ re-enter the horizon ($z \sim 10^3, 10^6$), the DM is effectively cold and its temperature makes negligible contribution to $R_\chi$.

\begin{figure}[!ht]
\includegraphics[width=\linewidth]{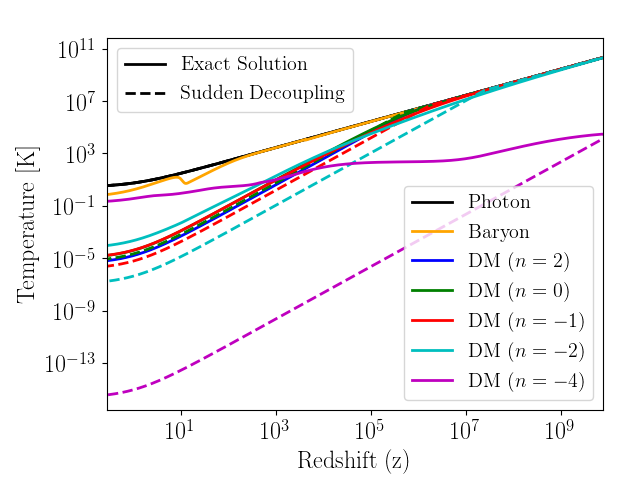}
\caption{Temperature evolution of dark matter, photons, and baryons evolved exactly with  Eqs. \ref{eq:tempev} and \ref{eq:barytempev} (solid lines) and using a sudden decoupling model as in Eq. \ref{eq:decoupling} (dashed lines). The DM mass is fixed at $m_\chi = 1$ GeV and $\sigma_0$ is fixed at the 95\% confidence level values from the last column of Table~\ref{sigmas} (CMB TTEE + Lyman-$\alpha$).  The DM temperature evolution after decoupling is approximated by $a^{-2}$ for $n>-4$. For $n=-4$ the DM temperature is negligible compared to the baryons, at the relevant redshifts for the data considered ($z\approx10^3$ and $z\approx 10^6$).}
\label{temp}
\end{figure}

The numerically-solved temperature evolution of dark matter (solid lines), along with our decoupling approximation (dashed lines) are shown in Figure~\ref{temp} for various choices of $n$, for a fixed $m_\chi = 1$ GeV,  using the 95\% CL values of $\sigma_0$ that come as a result of our analysis, quoted in the last column of Table \ref{sigmas} (CMB TTEE + Lyman-$\alpha$), and the $\Lambda$CDM cosmological parameters fixed to their no-scattering best fit values. As shown, a sudden decoupling model is more accurate for more positive $n$ scenarios. However, for all solutions the dark matter is cold compared to baryons at $z\sim 10^3$, so the CMB (which most strongly constrains the $n\leq -2$ scenarios) is insensitive to DM temperature of this amplitude.

In all cases, the relative difference between the exact and approximate temperature evolutions induced in the temperature and polarization power spectra produce a negligible likelihood difference.

\section{Numerical Results \label{NR}}

We modify the Boltzmann solver CAMB \cite{Lewis:2002ah} to include DM-proton elastic scattering and run a Markov Chain Monte Carlo (MCMC) likelihood analysis using CMB data (both temperature and polarization power spectra) from the Planck 2015 data release  \cite{Ade:2015xua} and measurements of Lyman-$\alpha$  flux power spectrum from the Sloan Digital Sky Survey (SDSS) \cite{McDonald:2004eu}. 

 The cosmological parameters varied in our Markov chains are the scattering amplitude $\sigma_0$ along with the standard $\Lambda$CDM parameters: the baryon density, $\Omega_b h^2$, the DM density, $\Omega_\chi h^2$, the optical depth to reionization, $\tau$, the angular size of the horizon at the time of recombination, $\theta_s$, and the amplitude and the tilt of the scalar perturbations, $\ln A_s$ and  $n_s$.  The power-law index $n$ and the DM particle mass $m_\chi$ are fixed within each MCMC run, and runs for  $m_\chi = 10$ GeV, 1 GeV, and 10 MeV are completed for each  $n \in \{ -4, -2, -1, 0 ,2\}$. We use the Gelman-Rubin criterion for convergence, and require that the ratio of variance between chains to the variance of an individual chain is less than $ 0.01$.

\begin{figure*}[!htb]
\centering
    \begin{subfigure}[t]{0.49\textwidth}
        \includegraphics[width=\linewidth]{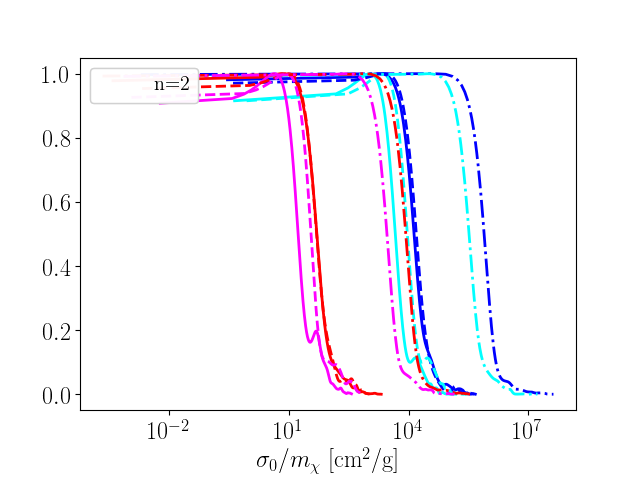}
    \end{subfigure}
    \begin{subfigure}[b]{0.49\textwidth}
        \includegraphics[width=\linewidth]{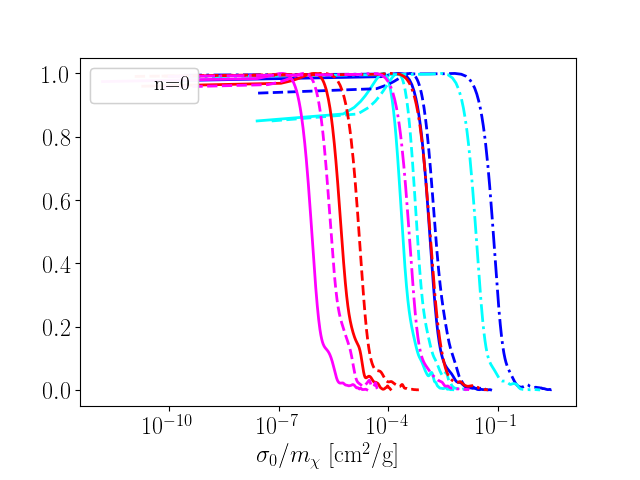}
    \end{subfigure}
~
    \begin{subfigure}[t]{0.49\textwidth}
        \includegraphics[width=\linewidth]{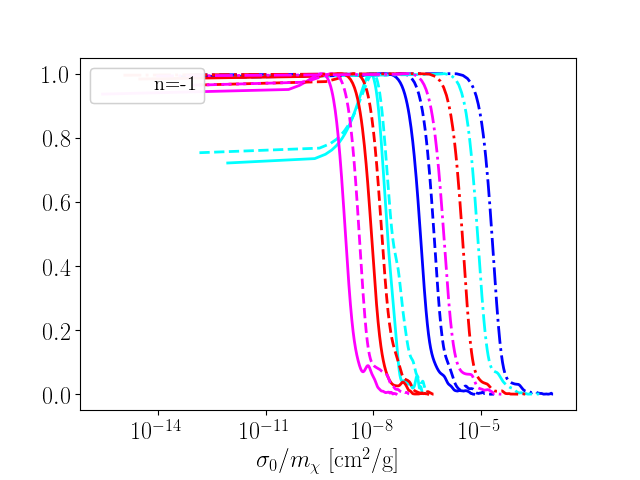}
    \end{subfigure}
    \begin{subfigure}[b]{0.49\textwidth}
        \includegraphics[width=\linewidth]{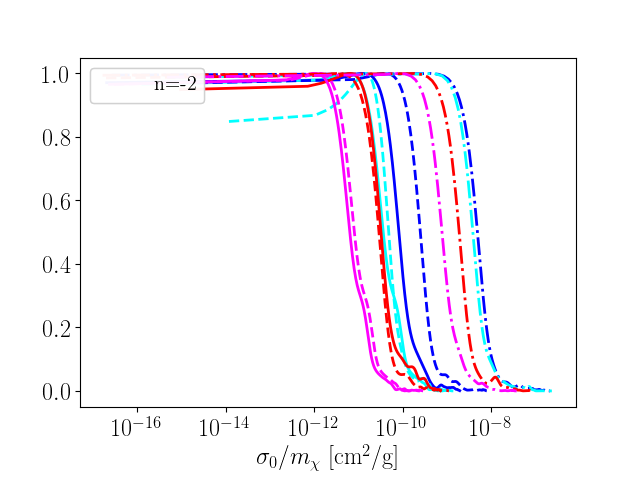}
    \end{subfigure}
~
    \begin{subfigure}[b]{0.49\textwidth}
        \includegraphics[width=\linewidth]{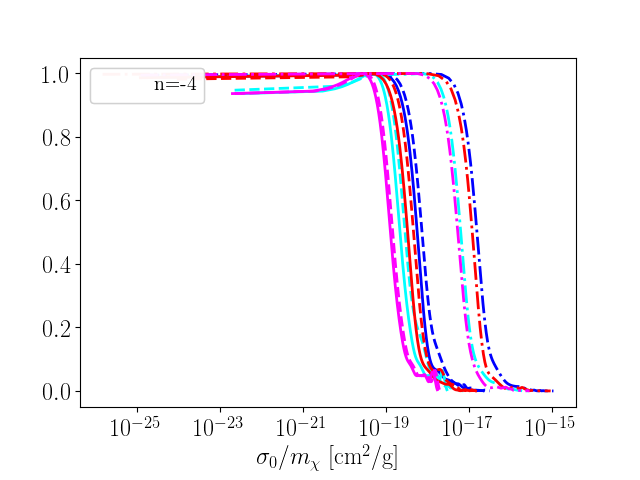}
    \end{subfigure}
    \begin{subfigure}[b]{0.49\textwidth}
\centering
        \includegraphics[width=0.5\linewidth]{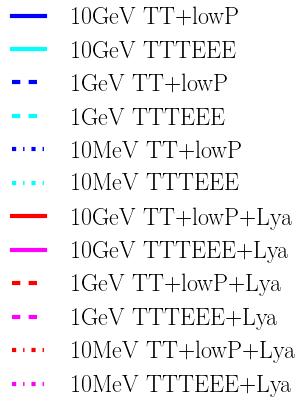}
\vspace{2mm}
    \end{subfigure}
\caption{1D posterior probability distribution functions for $\sigma_0/m_\chi$. The addition of Lyman-$\alpha$ forest data provides stronger constraints for scenarios with increasing (positive) values of $n$, whereas the inclusion of CMB data provides more stringent constraints for the more negative-$n$ scenarios.}
\label{sigmapdf}
\end{figure*}

Our  95\% confidence level limits on the upper-bound values of $\sigma_0$ for all values of $m_\chi$ are shown in Table \ref{sigmas}. As shown, scenarios with increasingly positive values of $n$ induce increasing amounts of suppression on small-scale structure, and thus can be better constrained by LSS data.  

\begin{figure}[!htb]
\centering
    \begin{subfigure}[t]{0.49\textwidth}
        \includegraphics[width=\linewidth]{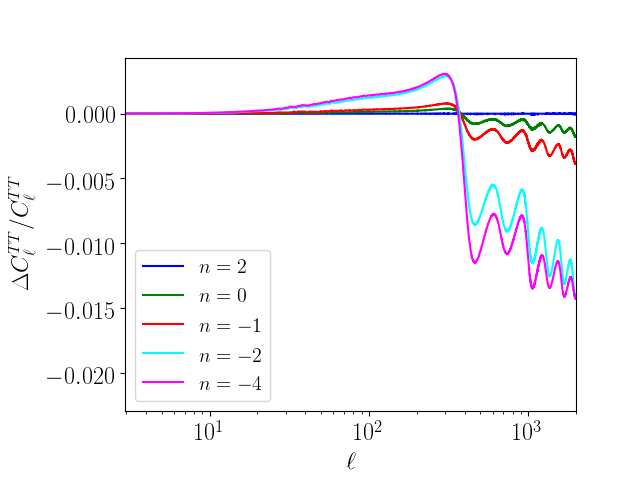}
    \end{subfigure}
~
    \begin{subfigure}[b]{0.49\textwidth}
        \includegraphics[width=\linewidth]{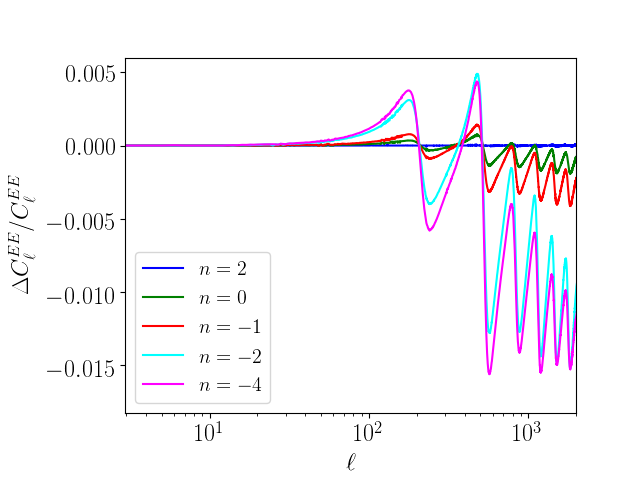}
    \end{subfigure}
\caption{Fractional difference in the CMB temperature (above) and $E$-mode polarization (below) power spectra of each $n$-scenario relative to the fiducial no-scattering case. Here, we fix $m_\chi =1$ GeV and take $\sigma_0$ to be the 95\% CL upper bounds in the last column of Table~\ref{sigmas} (CMB TTEE + Lyman-$\alpha$), with the remaining parameters fixed to the no-scattering best fit values.}
\label{cls1GeV}
\end{figure}

\begin{figure}[!t]
\centering
    \begin{subfigure}[t]{0.49\textwidth}
\includegraphics[width=\linewidth]{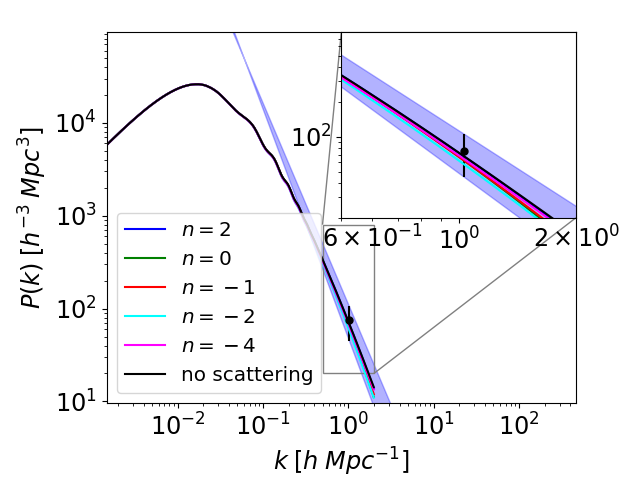}
    \end{subfigure}
\caption{Matter power spectrum for various $n$-scenarios and the fiducial no-scattering case. Here, we fix $m_\chi =1$ GeV and take $\sigma_0$ to be the $95 \%$ CL upper bounds from the last column of Table~\ref{sigmas} (CMB TTEE + Lyman-$\alpha$),  with the remaining parameters fixed to the no-scattering best fit values. The data point and violet band represent the amplitude and tilt, and respective 95\% CL error bars, derived from Lyman-$\alpha$ data.  The values are quoted from Ref. \cite{McDonald:2004eu}.}
\label{pk}
\end{figure}

The 1D posterior probability distributions of these various cases are shown in Figure~\ref{sigmapdf}. As can be seen from Table~\ref{sigmas}, the polarization power spectra are most sensitive to the $n=-4$ models; on the other hand, Lyman-$\alpha$ constrains most strongly models with positive $n$.
In Figure~\ref{cls1GeV} we show the fractional difference of the temperature and polarization CMB power spectra in models with scattering, relative to a fiducial zero-scattering cosmology. Figure~\ref{pk} similarly compares the matter power spectra generated by  various DM-baryon scattering scenarios. In both Figures~\ref{cls1GeV} and \ref{pk} the values used for $\sigma_0$ are the 95\% confidence level upper bounds from the last column in Table~\ref{sigmas} (CMB TTEE + Lyman-$\alpha$) where $m_\chi =1$ GeV, and the rest of the cosmological parameters are fixed to the no-scattering best fit values.    

\begin{table*}
\begin{center} 
\textbf{\large{$\sigma_0$ [cm$^2$]} ($m_\chi=$ 10 GeV)}

\vspace{1mm}
\begin{tabular} {c c| c| c| c c }
\hline \hline
\\
$n$  & CMB (TT + lowP) & CMB  (TT + lowP) + Ly-$\alpha$ & CMB (TTEE) & CMB  (TTEE) + Ly-$\alpha$\\ 
\hline
-4 & $2.1\times 10^{-40}$  &  $2.0 \times 10^{-40}$ & $8.6 \times 10^{-41}$ & $8.0 \times 10^{-41}$\\
-2 & $5.2 \times 10^{-32}$  & $1.0 \times 10^{-32}$ & $3.5 \times 10^{-32}$ & $9.2 \times 10^{-33}$\\
-1 & $2.9 \times 10^{-28}$  & $2.5 \times 10^{-29}$ & $2.0\times 10^{- 28}$ & $2.0 \times 10^{-29}$\\
0  & $2.5 \times 10^{-24}$ & $6.2 \times 10^{-26}$ & $1.9 \times 10^{-24}$ & $5.8 \times 10^{-26}$ \\
2  &  $2.7 \times 10^{-18}$  & $3.4 \times 10^{-20}$ & $2.0 \times 10^{-18}$ & $2.4 \times 10^{-20}$ \\
\end{tabular}

\vspace{5mm}

\textbf{\large{$\sigma_0$ [cm$^2$]} ($m_\chi=$ 1 GeV)}

\vspace{1mm}
\begin{tabular} {c c| c| c|  c c}
\hline \hline
\\
$n$  & CMB (TT + lowP) & CMB  (TT + lowP) + Ly-$\alpha$ & CMB (TTEE) & CMB  (TTEE) + Ly-$\alpha$\\ 
\hline
-4 & $4.3 \times 10^{-41}$  &  $4.1 \times 10^{-41}$ & $1.8 \times 10^{-41}$ & $1.6 \times 10^{-41}$\\
-2 & $1.0 \times 10^{-32}$  & $2.2 \times 10^{-33}$ & $6.8 \times 10^{-33}$ & $1.7 \times 10^{-33}$\\
-1 & $5.9 \times 10^{-29}$  & $5.0 \times 10^{-30}$ & $3.8 \times 10^{-29}$ & $3.6 \times 10^{-30}$\\
0  & $5.1 \times 10^{-25}$ & $1.3 \times 10^{-26}$ & $3.9 \times 10^{-25}$ & $1.2 \times 10^{-26}$ \\
2  &  $5.4 \times 10^{-19}$  & $6.8 \times 10^{-21}$ & $4.1 \times 10^{-19}$ & $4.9 \times 10^{-21}$ \\
\end{tabular}

\vspace{5mm}

\textbf{\large{$\sigma_0$ [cm$^2$]} ($m_\chi=$ 10 MeV)}

\vspace{1mm}
\begin{tabular} {c c |c |c  |c c}
\hline \hline
\\
$n$  & CMB (TT + lowP) & CMB  (TT + lowP) + Ly-$\alpha$ & CMB (TTEE) & CMB  (TTEE) + Ly-$\alpha$\\ 
\hline
-4 & $2.2 \times 10^{-41}$  &  $2.2 \times 10^{-41}$ & $9.6 \times 10^{-42}$ & $9.0 \times 10^{-42}$\\
-2 & $5.1 \times 10^{-33}$  & $1.1 \times 10^{-33}$ & $3.4\times 10^{-33}$ & $9.2 \times 10^{-34}$\\
-1 & $2.8 \times 10^{-29}$  & $2.5 \times 10^{-30}$ & $1.9 \times 10^{-29}$ & $1.8 \times 10^{-30}$\\
0  & $2.6 \times 10^{-25}$ & $6.4 \times 10^{-27}$ & $1.8 \times 10^{-25}$ & $5.6 \times 10^{-27}$ \\
2  &  $2.5 \times 10^{-19}$  & $3.3 \times 10^{-21}$ & $1.9 \times 10^{-19}$ & $2.3 \times 10^{-21}$ \\
\end{tabular}

\end{center}
\caption{95\% confidence level upper-bounds on $\sigma_0$ (in units of cm$^2$) from MCMC analyses  with various data sets. DM particle masses of $10$ GeV, 1 GeV, and 10 MeV are shown here for each choice of power-law scattering index. ``TT+ lowP" refers to the high-$\ell$ and low-$\ell$ CMB temperature and low-$\ell$ LFI polarization data, and ``TTTEEE" refers to the complete set of temperature and polarization data provided by the Planck 2015 data release.  ``Ly-$\alpha$" refers to the Lyman-$\alpha$ flux power spectrum data from the SDSS.}
\label{sigmas}
\end{table*}

\section{CMB Polarization sensitivity to DM-Baryon Scattering \label{PL}}

The addition of high-$\ell$ CMB polarization data provides a larger improvement on the the constraints for the $n=-4$ scenario, relative to the other n-scenarios considered in this work. This is because the CMB E-mode polarization is directly sourced by the velocity of the baryon-photon fluid, and it is therefore more sensitive to DM-baryon scattering.  

The source functions of CMB temperature and polarization fluctuations are given respectively by \cite{Zaldarriaga:1996xe} 

\begin{align}
S_T (k, \eta) = & g(\eta) [\Theta_0 + \Psi]  \nonumber\\ 
 +& \frac{d}{d\eta} \left( \frac{i v_b(k,\eta) g(\eta)}{k}\right)  + e^{-\tau} [\dot \Psi  - \dot \Phi ]
\end{align}
\begin{equation}
S_P (k, \mu, \eta) = g(\eta) \frac{3}{4} (1-\mu^2) (\Theta_2  + \Theta_{P0} + \Theta_{P2}),
\end{equation}
where the $\mu$ term encodes the on-sky geometry, $g(\eta)$ is the visibility function, $\Theta_{(P) \ell} (k, \eta)$ is the power of the temperature (polarization) $\ell$-th multipole of Fourier mode $k$ at conformal time $\eta$, and $\Phi$ and $\Psi$ parametrize the scalar metric perturbations. Overdots denote derivatives with respect to conformal time.

\begin{figure}[!htb]
\centering
\includegraphics[width=\linewidth]{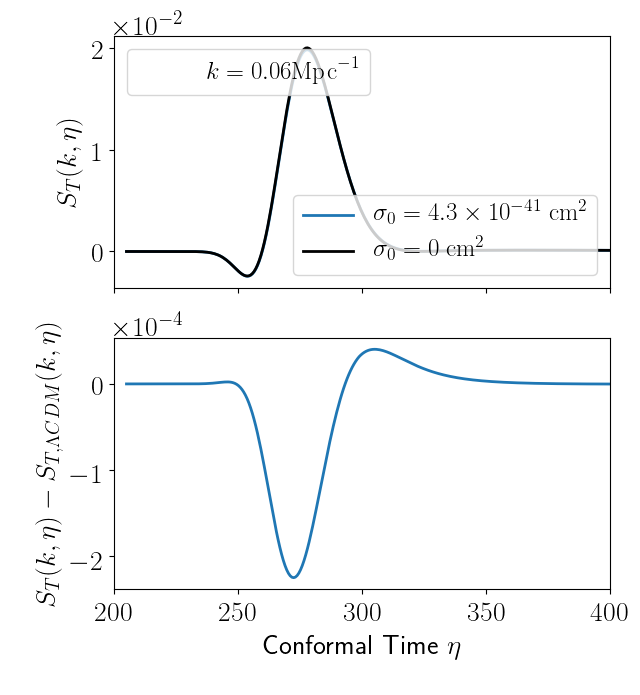}
\caption{The temperature anisotropy source function for the scattering cross section corresponding to the 95\% CL constraints derived from CMB TT+lowP data and the no scattering case, and their relative difference. We have restricted to the $n=-4$ scenario and taken $m_\chi = 1$ GeV.  As shown, the addition of DM-baryon interactions changes the source function by order $1\%$.}
\label{st}
\end{figure}

\begin{figure}[!htb]
\centering
\includegraphics[width=\linewidth]{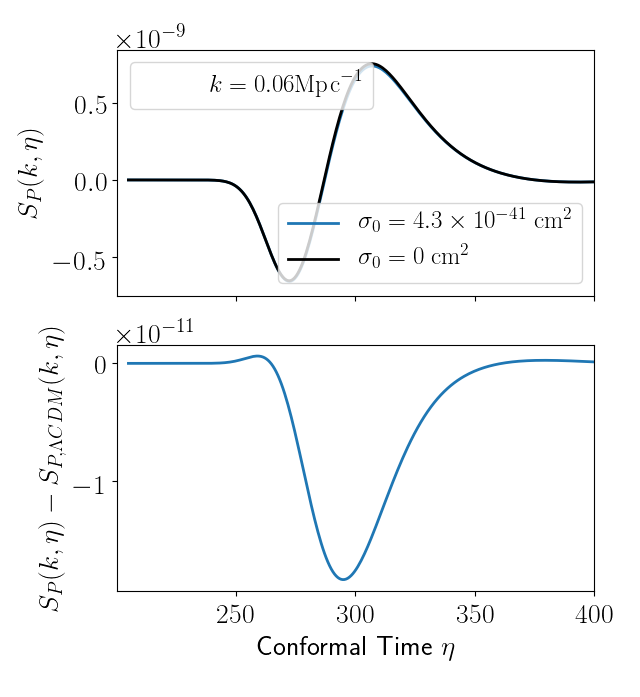}
\caption{Similar to Figure \ref{st}, but for the CMB E-mode polarization source function. As shown, the addition of DM-baryon elastic scattering suppresses the source amplitude by order $4\%$, showing a larger sensitivity of the polarization source relative to the temperature one.}
\label{sp}
\end{figure}

\begin{figure}[!htb]
\centering
\includegraphics[width=\linewidth]{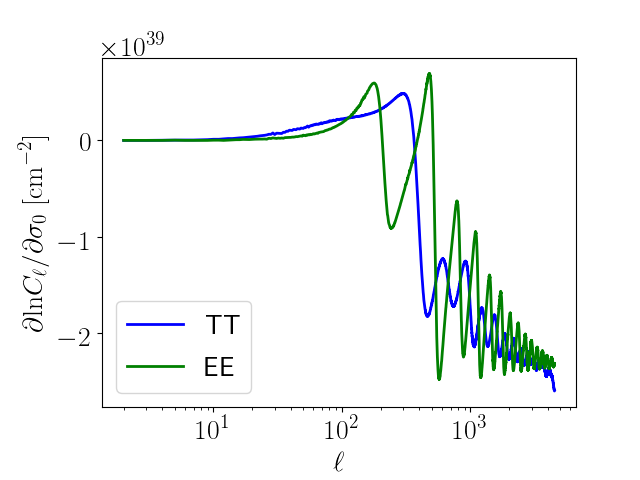}
\caption{The partial derivative of $\ln C_\ell$ with respect to DM-scattering cross-section $\sigma_0$. We have restricted to the $n=-4$ scenario and taken $m_\chi = 1$ GeV. The E-mode polarization power spectrum is shown to be a powerful tool for constraining DM-baryon interactions.}
\label{Clderiv}
\end{figure}

The temperature source function is dominated by the temperature monopole $\Theta_0$, whereas that of polarization is dominated by the much smaller temperature quadrupole $\Theta_2$.  Since the polarization source  is linearly dependent on the velocity of the baryon-photon fluid, turning on DM-baryon interactions results in a more significant change to the polarization source at every $k$-mode.  Figure~\ref{st} shows the amplitude of the temperature source at some arbitrary $k= 0.06\text{ Mpc}^{-1}$ ($\ell \approx 850$) and its difference to the no-scattering case. Figure~\ref{sp} shows the same for the polarization source. We can see that the polarization source exhibits a larger relative change upon allowing DM-baryon scattering. 
Figure~\ref{Clderiv} shows the derivative of both temperature and polarization spectra with respect to the DM-proton scattering cross-section, illustrating this difference. 

\section{Analytic Scaling of Constraints \label{MS}}

In this section, we propose a scaling of our MCMC constraints on $\sigma_0$ to apply to all $m_\chi \gtrsim$ 1 MeV.  The $\sigma_0 -m_\chi$ relation is set by two coefficients: the momentum exchange, given by $R_\chi$, defined in Eq. \ref{rchi}, and the thermal exchange rate, given by $m_\chi/(m_\chi + m_H) R_\chi$, as in Eqs.~\ref{eq:tempev} and \ref{eq:temperature_rate}. 

We assume that the dark matter scatters only with protons -- that is, we neglect DM-Helium and DM-electron scattering. We also assume non-relativistic kinematics at $z=10^9$, the starting point of our numerical analysis; thus, the maximal lower limit we can extend our results to is down to $m_\chi \sim $ 1 MeV. 

For effectively cold DM, $R_\chi$ can be approximated as being proportional to $ \sigma_0/ (m_\chi + m_H) $, if $T_\chi/m_\chi \ll T_H/m_H $ holds true.  This is because the baryon temperature is largely unaffected by elastic scattering with DM, for choices of cross section up to several orders of magnitude above our 95\% CL upper bound.   This reduces the momentum-based scaling and the temperature-based scaling to $\sigma_0 \propto (m_\chi + m_H)$ and $\sigma_0 \propto (m_\chi + m_H)^2/m_\chi$, respectively.

\begin{figure}[!t]
\centering
    \begin{subfigure}[t]{0.5\textwidth}
        \centering
        \includegraphics[width=\linewidth]{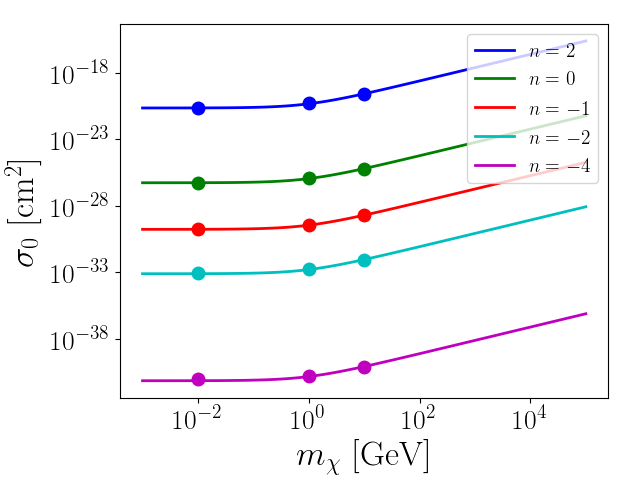}
    \end{subfigure}
\caption{Constraints for DM-baryon scattering at the 95\% CL in the $m_\chi -  \sigma_0$ parameter space from Planck temperature + polarization and Lyman-$\alpha$ forest data and our proposed extrapolation.}
\label{scaling}
\end{figure}

Fig. \ref{scaling} shows our 95\% CL exclusion constraints at $10$ GeV, $1$ GeV and $10$ MeV. After running our MCMC likelihood analysis, we find that the DM is sufficiently cold that the thermalization process is subdominant and the scaling relation is set almost entirely by the momentum exchange.  A momentum-based extrapolation from 1 GeV results is also shown to illustrate this.

We note that for $n \geq -1$, the scaling of constraints as $\sigma_0 \propto (m_\chi + m_H)$ is strictly conservative and valid up to the non-relativistic limit, since the temperature-dependent term in $R_\chi$, $(T_\chi/m_\chi + T_H/m_H)^{(n+1)/2}$, is given by a positive power-law. 

\begin{figure}[!htb]
\centering
    \begin{subfigure}[t]{0.5\textwidth}
        \centering
        \includegraphics[width=\linewidth]{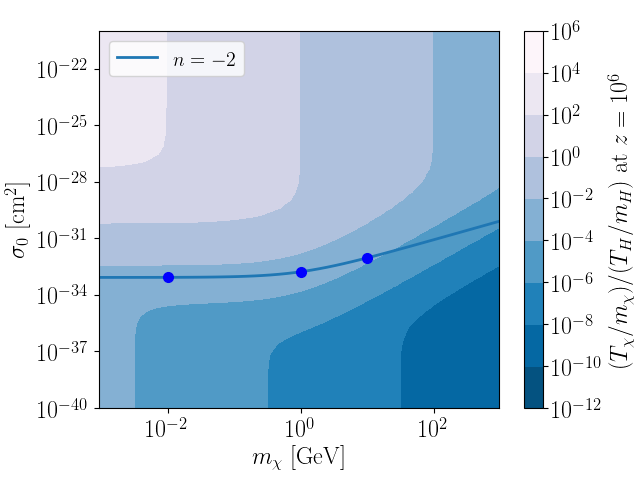}
	
    \end{subfigure}
\caption{Contours of $T_\chi m_H/(T_Hm_\chi)$ in the $\sigma_0-m_\chi$ plane for the $n=-2$ scenario, evaluated at $z=10^6$ (Lyman-$\alpha$ modes re-entry). For $T_\chi/m_\chi \ll T_H/m_H$, the scaling $\sigma_0 \propto (m_\chi + m_H)$ is valid (the solid curve represents this limit).  Data points (blue circles) are $95\%$ CL results from our MCMC likelihood analysis.}
\label{masslimnm2}
\end{figure}

\begin{figure}[!htb]
    \begin{subfigure}[t]{0.5\textwidth}
        \centering
        \includegraphics[width=\linewidth]{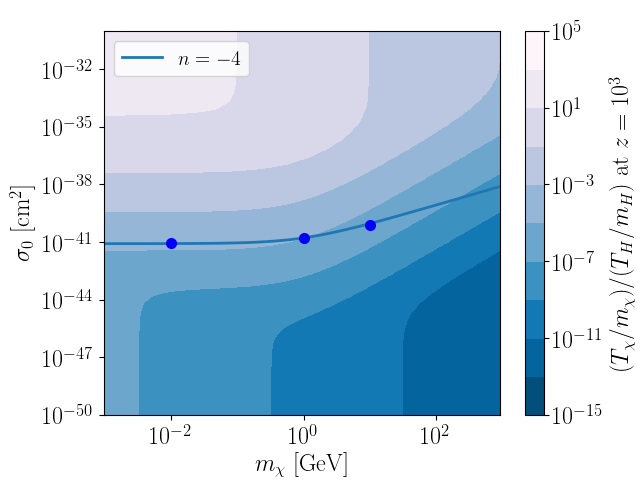}
	
    \end{subfigure}

\caption{ Similar to Fig. \ref{masslimnm2}, but for the $n=-4$ scenario, evaluated at $z=10^3$ (time of decoupling of the CMB).}
\label{masslimnm4}

\end{figure}

For $n \leq -2$ however, this approximation is not automatic: the temperature dependent term in $R_\chi$ carries a negative power index and a dominant $T_\chi/m_\chi$ term will suppress the scattering effect. Since $R_\chi$ is found to decrease with time for $n=-2$ and increase for $n=-4$, the former is predominantly constrained by Lyman-$\alpha$ data, whose modes re-enter the horizon at redshifts $z \simeq 10^6$, and the latter is predominantly constrained by CMB, with $z\simeq 10^3$ being the relevant redshift.  For $n=-4$ in particular the peculiar velocity term $V^2_{RMS}/3$ is important for redshifts $z<10^4$.  Figure ~\ref{masslimnm2} shows, for the $n=-2$ scenario, the region in $\sigma_0 - m_\chi$ parameter space where $T_\chi/m_\chi \ll T_H/m_H$  is valid at $z = 10^6$ ; Figure \ref{masslimnm4} does the same for $n=-4$ at $z = 10^3$. In these figures we also show our MCMC results at $m_\chi=10$ GeV, 1 GeV, and 10 MeV, as well as the extrapolation by $\sigma_0 \propto (m_\chi + m_H)$.  As shown, the proposed extension lies comfortably in the range of $T_\chi/m_\chi \ll T_H/m_H $ down to $m_\chi \approx$  1 MeV as well.

\begin{figure}[!t]
\centering
    \begin{subfigure}[t]{0.5\textwidth}
        \centering
        \includegraphics[width=\linewidth]{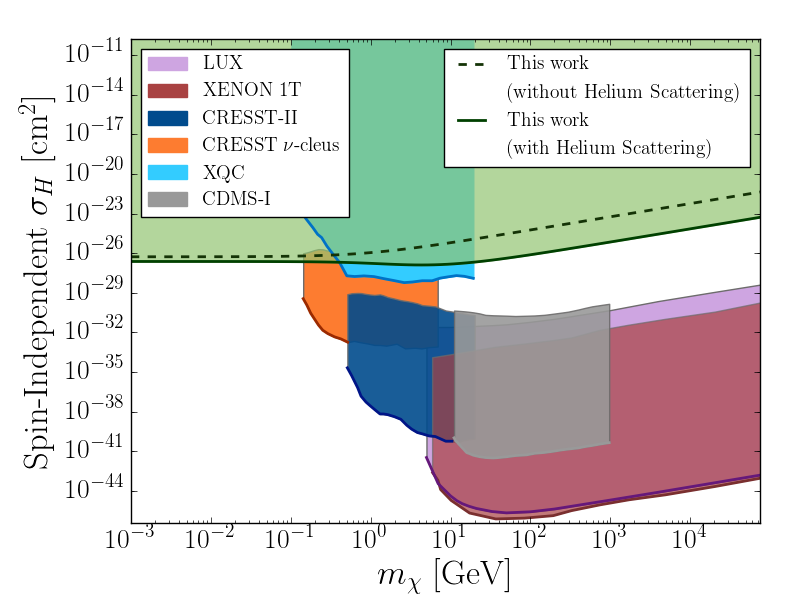}
    \end{subfigure}
\caption{Constraints on $n=0$ DM-baryon scattering in the $m_\chi - \sigma_H$ parameter space for underlying theory with (solid lines) and without (dashed lines) Helium scattering. Limits from direct detection searches are quoted from Refs. \cite{Akerib:2016vxi, Angloher:2015ewa, Angloher:2017sxg, Davis:2017noy, Mahdawi:2017cxz, Aprile:2017iyp, Kavanagh:2017cru}.} 
\label{directdetect}
\end{figure}

\section{Case Study: velocity and spin-independent scattering \label{DD}}

In this section we apply our results to the specific case of spin-independent $n=0$ elastic scattering, a particularly well-motivated effective interaction (cf.  for instance \cite{Chen:2002yh, Boehm:2001hm, Gluscevic:2017ywp, Fitzpatrick:2012ix} ) and probed extensively in nuclear-recoil type experiments. 

Since specializing to this model allows us to write down the DM-Helium scattering cross-section $\sigma_{He}$ as a specific function of the DM-proton cross-section, we can extend our previous results to account for DM-Helium interactions as well. $R_\chi$ is now an effective momentum-transfer rate that encompasses both  DM-proton and DM-Helium momentum transfer: $R_{\chi} = R_{\chi,p} + R_{\chi, He}$, where, in the $n=0$ case,

\begin{equation}
 R_{\chi, i } = \frac{a c_0 \rho_i \sigma_{i}}{m_\chi + m_i} v_{\chi,i}.
\end{equation}

Here, $c_0$ is a numerical factor shown in Table ~\ref{cns} in the Appendix, and $v_{\chi,i }$ is the relative velocity of DM and particle species $i$, that can be either unbound protons or Helium.

Following the treatment of  Refs. \cite{Catena:2015uha, Gluscevic:2017ywp}, we can write the DM-Helium momentum transfer rate as

\begin{eqnarray}
R_{\chi,He} &= & \; {a c_0 \rho_{He}\over m_\chi + m_{He}}\sigma_{He}  v_{\chi, He} \left( 1 + (2 \mu_{\chi He} a_{He} v_{\chi, He} )^2 \right)^{-2}\nonumber \\
& \simeq& {a c_0 \rho_{He}\over m_\chi + m_{He}}\sigma_{He}  v_{\chi, He},
  \end{eqnarray}
and
  \begin{equation}
\sigma_{He}=  \; 4  \frac{\mu^2_{\chi He} }{\mu^2_{\chi H}}\sigma_H.
 \end{equation}
 
Here, $\mu_{\chi i} =  m_\chi m_i/(m_\chi + m_i)$ is the reduced mass of the DM-$i$ system, and $a_He \simeq 1.5$ fm is nuclear shell length parameter for Helium \cite{Fitzpatrick:2012ix, Gluscevic:2017ywp}.  The simplification in the second line is based on the assumption that we are in the non-relativistic regime, $v_{\chi, He} \ll 1$.  Similarly, we assume that all baryons share the same temperature and peculiar velocity relative to DM, and use $v_{\chi He} \gtrsim \frac{1}{2} v_{\chi p}$.  The total momentum-transfer is then 

\begin{align}
R_\chi &  = \frac{a c_0  \rho_b v_{\chi, H} \mathcal{F}_{He} }{m_\chi + m_H} \sigma_0  \nonumber\\
& \gtrsim  a c_0  n_b v_{\chi, H} \left( \frac{m_H \sigma_H \mathcal{F}_{He}}{m_\chi + m_H}  + \frac{m_{He} \sigma_{He} (1- \mathcal{F}_{He})}{2( m_\chi + m_{He})}  \right)   \nonumber \\
& \simeq \frac{a c_0  \rho_b v_{\chi, H} \mathcal{F}_{He} }{m_\chi + m_H}  \sigma_H   \left( 1 + \frac{1- \mathcal{F}_{He}}{\mathcal{F}_{He}} \frac{ 2\mu^3_{\chi He}}{\mu^3_{\chi H} } \right).
\end{align}

This provides a straightforward, albeit conservative, relation between our numerical variable $\sigma_0$ and the ``Helium-subtracted" cross-section $\sigma_H$ in the case of spin-independent $n=0$ scattering. This improves our results by as much as a factor of 20 in the high-mass regime.

Figure~\ref{directdetect} shows the regions we have excluded at the $2-\sigma$ level in the $m_\chi - \sigma_H$ parameter space compared to regions explored by direct detection experiments XENON-1T \cite{Aprile:2017iyp}, LUX \cite{Akerib:2016vxi}, XQC \cite{Mahdawi:2017cxz, Erickcek:2007jv}, CRESST-II\cite{Angloher:2015ewa}, the CRESST $\nu$-cleus Surface Run \cite{Angloher:2017sxg, Davis:2017noy} , and the CDMS-I re-analysis \cite{Kavanagh:2017cru}.  While nuclear recoil experiments provide high sensitivity at high masses, direct detection limits towards sub-GeV dark matter are currently restricted to DM-electron scattering,  \cite{Essig:2015cda,Tiffenberg:2017aac,Essig:2012yx},  and sensitivity of underground experiments in particular are cut off at high cross-sections by scattering through the rock overburden \cite{Kouvaris:2014lpa, Davis:2017noy}. Cosmological observables are thus especially complementary in this regime.

\begin{figure}[!t]
\centering
    \begin{subfigure}[t]{0.5\textwidth}
        \centering
        \includegraphics[width=\linewidth]{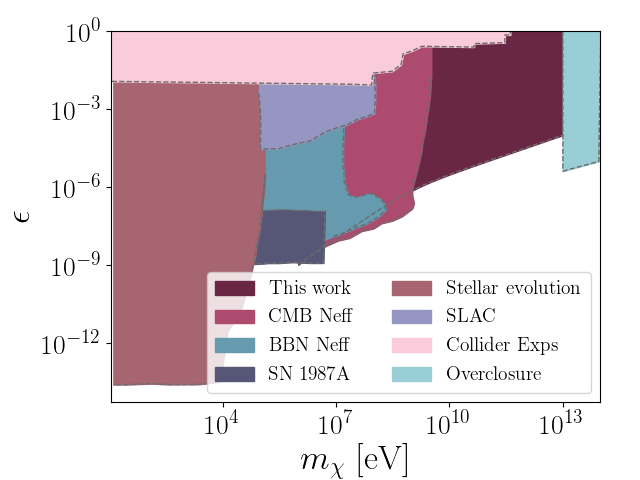}
    \end{subfigure}
\caption{Constraints from this work on millicharged DM scattering (corresponding to the $n=-4$ scenario)  in $\epsilon-m_\chi$ space compared to bounds from other areas: cooling of giants, white dwarfs, and supernovae and constraints on $N_{eff}$ from BBN and CMB \cite{Vinyoles:2015khy, Vogel:2013raa}, overclosure of the Universe \cite{Davidson:1991si} and various collider experiments \cite{Jaeckel:2012yz, Davidson:2000hf, Prinz:1998ua, Davidson:1991si}. We have assumed here that all DM is millicharged.}
\label{millicharge}
\end{figure}

\section{Case Study: Millicharged DM \label{MC}}

We will now consider the scenario of millicharged DM, explored previously in Refs. \cite{Dolgov:2013una,  Dubovsky:2003yn, Davidson:2000hf, Cline:2012is, McDermott:2010pa, Vogel:2013raa,Agrawal:2017pnb}. For this case, we assume that all DM is charged under some hidden U(1) gauge with a ``dark photon", which kinetically mixes with the Standard Model photon such that DM particles carry a fractional electromagnetic charge $\epsilon e$.  The non-relativistic DM-proton scattering thus follows a Coulomb cross-section: 
\begin{equation} 
\frac{d\sigma}{d\Omega} = \frac{\epsilon^2 \alpha^2_{EM}}{4 \sin^4{\theta/2}} \mu_{\chi b}^{-2} v^{-4},
\end{equation}
 and we see that our $n=-4$ constraints are applicable here. 

To regulate the divergence at small scattering angles, we impose a minimum scattering angle $\theta_{min}$ set by the Debye screening length of the baryon plasma  
\begin{equation}  
\theta_{min} = \frac{2\epsilon \alpha_{EM}}{3 T \lambda_D}, \qquad \lambda_D = \sqrt{\frac{T}{4\pi \alpha_{EM} n_e}},
\end{equation}

such that we can apply our results 
\begin{equation}
\sigma(v) =2 \pi \int^\pi_{\theta_{min}} \left(1-\cos(\theta)\right)\dd \theta \sin \theta  \frac{d\sigma}{d\Omega}. 
\end{equation}

We obtain the approximate numerical bound \begin{equation} \epsilon <  1.0 \times 10^{-6} \left(\frac{m_\chi}{\text{GeV}} \right)^{1/2}  \left(\frac{ \mu_{\chi b}}{\text{GeV}} \right)^{1/2}. \end{equation}

Constraints on millicharged DM particles in the low-mass $\lesssim$ 1 MeV regime come predominantly from cooling dynamics of stars and supernovae, as well as constraints on the effective neutrino number $N_{eff}$ during Big Bang Nucleosynthesis (BBN) and CMB epochs \cite{Vogel:2013raa, Vinyoles:2015khy}.   Limits arise also from collider experiments such as from LHC and SLAC \cite{Jaeckel:2012yz, Davidson:2000hf, Prinz:1998ua, Davidson:1991si}. An additional constraint comes from rapid annihilation of high-mass DM inducing premature closure of the universe \cite{Davidson:1991si}.  Figure~\ref{millicharge} compares the bounds from this work with the previously mentioned results. As shown, CMB temperature and polarization data together with Lyman-$\alpha$ flux power spectrum measurements provide sensitive constraints to the scenario where all DM carries a millicharge.

\section{Conclusions}

In this work we consider a general class of elastic DM-proton interaction scenarios where the scattering cross-section scales phenomenologically as a power of relative velocity between protons and dark matter.  We perform an MCMC likelihood analysis and obtain constraints on the scattering cross section $\sigma_0$ for 10 GeV, 1 GeV, and 10 MeV dark matter particle masses and a range of power laws $n \in \{ -4, -2, -1, 0 ,2\}$, using CMB temperature and polarization data from the Planck satellite, and Lyman-$\alpha$ flux power spectrum data from the SDSS.  

We extend previous results with the addition of CMB polarization data, and find that it has a larger impact (relative to Lyman-$\alpha$) on scenarios with $n\leq-2$ because these scenarios are more sensitive to the evolution of perturbations at $z<10^4$. For positive-n scenarios, large-scale structure data remains the limiting source of constraint.  

Extrapolating our MCMC results to lower masses, we propose the scaling $\sigma_0 \propto (m_\chi + m_H)$, and show that this is valid until $m_\chi \approx$ 1 MeV, where the assumption of non-relativistic kinematics breaks down. This allows us to explore lower-mass regions of the $\sigma_0 - m_\chi$ parameter space, which are difficult to access with nuclear recoil experiments due to kinematic limitations.

Allowing for relativistic scattering dynamics will be necessary to extend this approach below the MeV scale. We leave this to future work.

\acknowledgments

It is our pleasure to thank Vera Gluscevic, Azadeh Moradinezhad Dizgah, Julian B.~Mu\~noz for helpful discussions.
We particularly thank Tracy Slatyer and Chih-Liang Wu for providing various cross-checks of our results.
This work was supported by the Dean's Competitive Fund for Promising Scholarship at Harvard University.

\medskip

\bibliography{Paper_Text}

\begin{appendix}
\section{Boltzmann Equations for DM-Baryon Scattering}\label{sec: AppendixA}

In this Appendix, we review the formulation of the modified Boltzmann equations in the presence of DM-baryon interactions, specifically with cross-sections that scale with relative DM-baryon velocity $v$ as $\sigma \propto  v^n$ for some index $n$. A more detailed treatment can be found in Ref. \cite{Dvorkin:2013cea}. 

We assume non-relativistic kinematics for both DM and baryons, which is accurate for $m_\chi $ above the MeV scale and $z\lesssim 10^9$

\subsection{Dark Matter - Baryon Drag Force}

Here we review the modifications to the standard Boltzmann equations derived in Ref. \cite{Dvorkin:2013cea}. For baryons and DM we assume a Maxwell distribution for their velocity distributions in the early Universe, 

\begin{eqnarray} \label{fvs} 
f_b(v_b) &=& \sqrt{\frac{2 m_b^3}{\pi T_b^3}} \exp \left[- \frac{v_b^2}{2(T_b/m_b)^2}\right] \\ \label{fvs2} f_\chi(v_\chi) &=& \sqrt{\frac{2 m_\chi^3}{\pi T_\chi^3}} \exp \left[ -\frac{(\vec v_\chi - \vec V_\chi)^2}{ 2(T_\chi / m_\chi)^2}\right],
\end{eqnarray}
where we take the baryon distribution to be isotropic and the DM population to be boosted with peculiar velocity $\vec V_\chi$ relative to this frame.  The baryon particle mass $m_b$ is taken to be the proton mass. Elastic collisions with the baryon fluid will eventually drive the DM population to isotropy. A given DM particle with velocity $v_\chi$ elastically colliding with a baryon of velocity $v_b$ experiences a change of momentum
\begin{equation} 
\Delta \vec p_\chi = \frac{m_\chi m_b}{m_\chi + m_b}  |\vec v_\chi - \vec v_b | \left(\hat n  -  \frac{\vec v_\chi - \vec v_b}{| \vec v_\chi - \vec v_b |} \right),
\end{equation} 
with $\hat n$ being the outgoing direction of the scattered DM particle.

Taking the momentum-transfer scattering cross section as
\begin{equation} 
\sigma (v) = \sigma_0 v^n,
\end{equation}
and integrating over the entire baryon fluid, the overall deceleration of the DM particle can be written as 
\begin{align} 
\label{eq::dVchi}
\frac{\dd \vec v_\chi}{\dd t}  =  - \frac{\rho_b \sigma_0}{m_\chi + m_b} & \int d^3\vec v_b f_b (v_b) \\  &\times |\vec v_\chi - \vec v_b|^{n+1} (\vec v_\chi - \vec v_b) \nonumber
\end{align}     
where $\rho_b$ is the baryon mass density.  The latter integral encodes the dependence on power-law index $n$. In turn, integrating over the DM velocity distribution, we obtain the induced deceleration of the peculiar velocity 
\begin{equation} 
\frac{\dd \vv V_\chi}{\dd t} = \int d^3\vec v_\chi  \frac{\dd \vv v_\chi}{\dd t} f_\chi (v_\chi).
\end{equation}

$\dd \vv V_\chi / \dd t$ is dominated by two velocity scales. The first is $\vv V_\chi$ itself, and the second is the averaged velocity dispersion 
\begin{equation} 
\langle |\Delta \vec v|^2 \rangle =  \langle |\vec v_\chi - \vec v_b|^2 \rangle = 3 \left( \frac{T_b}{m_b} +  \frac{T_\chi}{m_\chi} \right).
\end{equation}

In the early universe, when thermal dispersion dominates, the integral Eq.~\ref{eq::dVchi} gives
\begin{equation} 
\label{vper} \frac{\dd \vv V_\chi}{\dd t} = -\vv V_\chi \frac{\rho_b \sigma_0 c_n}{m_\chi + m_b}  \left(  \frac{\langle |\Delta \vv v|^2 \rangle}{3}  \right)^{(n+1)/2},
\end{equation} 
valid to leading order in $V_\chi^2 / \langle (\Delta \vv v)^2 \rangle$. The constants $c_n$ are computed for the values of $n$ of interest and tabulated below.

\begin{table} [!htb]
\begin{center}
\begin{tabular}{ | c | | c | c | c | c | c | c | c | } 
\hline
$n$   & -4 & -3 & -2 & -1 & 0 &1 &2 \\
\hline
\hline 
$c_n$  & 0.27 & 0.33 & 0.53 & 1 & 2.1 &  5 &  13\\
\hline
\end{tabular}
\end{center}
\caption{\small Integration constants $c_n$ for different values of $n$.}
\label{cns}
\end{table}

At later times (after $z\simeq 10^4$) the peculiar velocity dominates and the the integral Eq.~\ref{eq::dVchi} gives for the DM deceleration, to leading order
\begin{equation}
\frac{\dd \vv V_\chi}{\dd t} = -\vv V_\chi \frac{\rho_b \sigma_0}{m_\chi + m_b}  V_\chi ^{n+1}.
\end{equation} 
At this point, the dependence becomes non-linear (unless $n=-1$), and, following Ref. \cite{Dvorkin:2013cea}, we will include a mean-field term for peculiar velocity when calculating the momentum transfer (see Eq.~\ref{eq:vrms}). 

\subsection{Modified Boltzmann Equations}

In this subsection, we modify  Boltzmann equations to account for DM-baryon scattering. We will work in synchronous gauge, following formulations in Ref. \cite{Dvorkin:2013cea, Ma:1995}, but allowing for nonzero peculiar velocity in dark matter.  For a given Fourier mode $k$ the density fluctuations $\delta_\chi$ and $\delta_b$  and velocity divergences  $\theta_\chi$ and $\theta_b$ of the DM and baryon fluids obey the evolution equations presented in the main text,

\begin{eqnarray} 
\dot \delta_\chi &=& -\theta_\chi - \frac{\dot h}{2} \\ 
\dot \delta_b &=& -\theta_b- \frac{\dot h}{2}  \\
\dot \theta_\chi &=& -\frac{\dot a}{a}\theta_\chi  + c^2_\chi k^2 \delta_\chi + R_\chi (\theta_b - \theta_\chi ) \label{boltzmann1}\\
\dot\theta_b&=& -\frac{\dot a}{a}\theta_b + c^2_b k^2 \delta_b + R_\gamma (\theta_\gamma - \theta_b) \nonumber \\ \phantom{\frac{\dd \theta_b}{\dd \eta}} && \phantom{-\frac{\dot a}{a}\theta_b + c^2_b k^2 \delta_b} + \frac{\rho_\chi}{\rho_b} R_\chi (\theta_\chi- \theta_b),
\label{boltzmann2}
\end{eqnarray} 
where overdots denote derivatives with respect to conformal time, $h$ is the metric perturbation, $c_\chi$ and $c_b$ refer respectively to the DM and baryon sound speeds, $R_\chi$ is the momentum-transfer coefficient for DM-baryon coupling,  and $R_\gamma$ is the coefficient for baryon-photon coupling (Ref.~\cite{Ma:1995}),
\begin{equation}
 R_\gamma = \frac{4\rho_\gamma}{3\rho_b} a n_e \sigma_T,
\end{equation}
where $\rho_\gamma$ is the photon energy density, $n_e$ is the electron number density, and $\sigma_T$ is the Thomson cross-section. 

The DM-baryon coupling term arises from the deceleration of the DM bulk velocity, given to leading order by Eq. \ref{vper}  in the limit of $V_\chi \ll \langle |\Delta \vv v|^2 \rangle$,
\begin{equation} 
R_\chi =  \frac{a \rho_b \sigma_0 c_n }{m_\chi + m_b} \left(  \frac{T_b}{m_b} + \frac{T_\chi}{m_\chi}\right)^{(n+1)/2} \mathcal{F}_{He},
\end{equation} 
and the corresponding factor contributing to $\dot \theta_b$ is weighted by the DM mass density.

The above equation is valid strictly for the $z>10^4$ regime, when the thermal velocity dispersion dominates over the DM bulk velocity (see Ref. \cite{Dvorkin:2013cea}). In order to extend the validity of our approach  beyond $z\simeq 10^4$, we add in by hand the averaged value of $V^2_\chi$, 
\begin{equation} 
V^2_{RMS} \equiv \langle V^2_\chi \rangle 
\simeq  \begin{cases} 10^{-8}, & z>10^3 \\ 10^{-8} \left(\frac{(1+z)}{10^3}\right)^2,  & z\leq 10^3, \end{cases} 
\end{equation}
to approximate $R_\chi$ at late times, where the thermal velocity is no longer dominant.  The modified momentum-exchange coefficient is then 
\begin{equation} 
R_\chi =  \frac{a \rho_b \sigma_0 c_n }{m_\chi + m_b} \left(  \frac{T_b}{m_b} + \frac{T_\chi}{m_\chi} + \frac{V^2_{RMS}}{3}\right)^{\frac{n+1}{2}} \mathcal{F}_{He}. 
 \end{equation}

The factor $\mathcal{F}_{He}$ is a corrective factor to account for the Helium fraction in baryons, and encodes dynamics for DM scattering off of Helium.  Assuming the baryons share a temperature and have no relative bulk velocity between species, this is given by 

\begin{align}
\mathcal{F}_{He} = & 1 - Y_{He} + Y_{He} \frac{\sigma_{He}}{\sigma_H} \frac{m_\chi + m_H}{m_\chi + m_{He}}  \nonumber \\
& \times \left( \frac{ \frac{T_\chi}{m_\chi} + \frac{T_b}{m_H} + V^2_{RMS}}{\frac{T_\chi}{m_\chi} + \frac{T_b}{m_{He}} + V^2_{RMS}} \right)^{\frac{n+1}{2}},
\end{align}
where $Y_{He}\approx 0.24$. For this work we conservatively assume that $\mathcal{F}_{He} \approx 0.76$.

The DM and baryon fluid temperatures evolve as 
\begin{eqnarray}  \dot T_\chi  &=& -2 \frac{\dot a}{a} T_\chi   + \frac{2 m_\chi}{m_\chi + m_b} R'_\chi (T_b - T_\chi) \\ \dot T_b &=& -2 \frac{\dot a}{a} T_b + \frac{2 \mu_b}{m_e} R'_\gamma (T_\gamma - T_b) \nonumber \\  & & \phantom{-2 \frac{\dot a}{a} T_b} + \frac{2 \mu_b}{m_\chi + m_b} \frac{\rho_\chi}{\rho_b}R'_\chi (T_\chi - T_b),
\end{eqnarray} 
where the non-adiabatic terms are due to DM-baryon scattering (thermalization rate $R'_\chi$) and photon-baryon coupling (Compton term $R'_\gamma$).  
Here, $\mu_b \simeq m_b (n_H + 4 n_{He})/(n_H + n_e + n_{He})$ is the baryon mean molecular weight.

To derive the DM-baryon thermalization rate $R'_\chi$, note that the change in DM energy upon nonrelativistic collision with a baryon is  $\Delta \epsilon_\chi = \Delta \vec p _\chi \cdot \vec v$, where $\vec v$ is the center-of-mass velocity. The specific heating rate of DM can then be found by integrating over the Maxwellian distributions of baryon and DM velocities in  Eqs.~\ref{fvs} - \ref{fvs2},  
\begin{align}  
\frac{\dd Q_\chi}{\dd t} &= - \frac{m_\chi \rho_b \sigma_0}{(m_\chi + m_b)^2} \int d^3 \vec v_\chi f_\chi(v_\chi) \\
&\times \int d^3 \vec v_b f_b(v_b) |\vec v_\chi - \vec v_b|^{n+1} (m_\chi \vec v_\chi - m_b \vec v_b) \cdot (\vec v_\chi - \vec v_b).  \nonumber
\end{align}

Integrating similarly to Eq. \ref{vper}, restricting to specific case of $\sigma_{He}=0$. and taking once more the limit of low peculiar velocity, 
\begin{equation}
\frac{\dd Q_\chi}{\dd t}  = -\frac{3ac_n m_\chi \rho_b \sigma_0}{(m_\chi + m_b)^2} \left(  \frac{T_b}{m_b} + \frac{T_\chi}{m_\chi}  \right) ^{\frac{n+1}{2}} (T_\chi - T_b) \label{dQdt}.
\end{equation}

Taking the DM fluid as an ideal gas $Q_\chi = 3T_\chi/2$, and adding in the corrective factors for helium fraction and $V_{\rm RMS}$ as before, we obtain the contribution on DM temperature evolution made by DM-baryon scattering,
\begin{eqnarray} 
\dot T_{\chi, b\chi} &=& - \frac{2a c_n m_\chi \rho_b \sigma_0}{(m_\chi + m_b)^2}  \mathcal{F}_{He}  (T_\chi - T_b)\nonumber \\ && \times \left(  \frac{T_b}{m_b} + \frac{T_\chi}{m_\chi}  + \frac{V^2_{RMS}}{3} \right) ^{(n+1)/2}\nonumber \\  &\equiv& \frac{2 m_\chi}{m_\chi + m_b}  R'_\chi   (T_b - T_\chi) \label{eq:temperature_rate}
\end{eqnarray} 
and thus the thermalization rate $R'_\chi$, equal to the momentum-exchange rate $R_\chi$ for $\sigma_{He}=0$. Note the corresponding baryon temperature term is weighted relative to the DM term by both $\mu_b/m_\chi$ and $\rho_\chi/\rho_b$. 

\subsection{Tight Coupling Approximation with DM-baryon drag}

Following Refs.~\cite{Sigurdson:2004zp,Ma:1995}, we derive equations for evolving the coupled DM, baryon, and photon fluids through the era of tight coupling, when the photon scattering rate $\tau_c^{-1} >> \dot{a}/{a}$.  We first rewrite the baryon evolution equation equation given in Eqs. ~\ref{boltzmann1} - \ref{boltzmann2} in terms of characteristic time scales:

\begin{align}
\label{ex}
\dot \theta_b&=-\frac{\dot a}{a}\theta_b+c_b^2k^2\delta_b+\frac{R}{\tc}(\theta_\gamma-\theta_b)+\frac{S}{\tb}(\theta_\chi-\theta_b).
\end{align}

We define $R$ (not to be confused with $R_\gamma$ or $R_\chi$) as $R=\frac{4\rho_\gamma}{3\rho_b} \propto a^{-1}$ and $S=\frac{\rho_\chi}{\rho_b} = \text{constant}$.  The conformal time scale of Thomson scattering is $\tc = (an_e\sigma_T)^{-1}$ is the conformal time scale of Thomson scattering, and similarly $\tb = R_\chi^{-1}$ gives the conformal time scale of the dark matter-baryon interaction. 

We will also need the photon velocity divergence equation (Ref.~\cite{Ma:1995}):
\begin{equation}
\label{photexact}
\dot \theta_\gamma = k^2\left(\frac{1}{4}\delta_\gamma-\sigma_\gamma\right)-\frac{1}{\tc}(\theta_\gamma-\theta_b)
\end{equation}

In the tight-coupling regime, $\tc$ is small compared to the conformal Hubble time, and the above differential equations become stiff. In order to solve these tightly coupled equations numerically, we find  equations for $\dot\theta_b$ (and consequently also for $ \dot \theta_\gamma$) in terms of the slip derivative $\dot \Theta_{\gamma \beta} = \dot \theta_\gamma- \dot \theta_b$, which we solve for in powers of $\tc$. 

Adding Eqs.~\ref{ex} and ~\ref{photexact}, and multiplying by $\tc$, gives an exact equation for the photon-baryon slip $\T = \theta_\gamma-\theta_b$,

\begin{eqnarray} 
\T &=& \frac{\tc}{1+R}\left[\dot -\Theta_{\gamma \beta} +\frac{\dot a}{a}\theta_b+k^2\left(\frac{1}{4}\delta_\gamma-c_b^2\delta_b-\sigma_\gamma\right)\right. \nonumber \\
&&\left.-\frac{S}{\tb}(\theta_\chi-\theta_b)\right]
\label{slipde}
\end{eqnarray} 

From Eq.~\ref{slipde}, we verify that the slip is first order in $\tc$. Differentiating, dropping terms of order $\tc^2$ (i.e $\ddot{\Theta}_{\gamma \beta}$) and using $\dot R= -\frac{\dot a}{a}R$ and $\dot S = 0$, we have
\begin{eqnarray} 
\dot \Theta_{\gamma \beta}  &=& \left(\frac{\dot \tau_c}{\tc}+\frac{R}{1+R}\frac{\dot a}{a}\right)\Theta_{\gamma\beta}+ \frac{\tc}{1+R} \nonumber\\
&\times&\left(-\dot X-\frac{S}{\tb}(\dot\theta_\chi-\dot\theta_b)+\frac{S\dot\tb}{\tb^2}(\theta_\chi-\theta_b)\right),
\label{eq1}
\end{eqnarray} 
where to first order in  $\tc$

\begin{align}
\label{eq2}
-\dot X =&\frac{\dot a}{a}\dot \theta_b + \frac{\ddot a}{a}\theta_b -\left(\frac{\dot a}{a}\right)^2 \theta_b  \nonumber \\  
& +k^2\left(\frac{1}{4}\dot \delta_\gamma -\dot \sigma_\gamma - c_b^2\dot \delta_b \right) \nonumber \\
=&2\frac{\dot a}{a}\dot \theta_b + \frac{\ddot a}{a}\theta_b \nonumber\\ 
& +k^2\left(\frac{1}{4}\dot \delta_\gamma - \frac{\dot a}{a}c_b^2 \delta_b - c_b^2\dot \delta_b - \dot \sigma_\gamma\right) \nonumber\\ 
&-\frac{R}{\tc}\frac{\dot a}{a}\T-\frac{S}{\tb}\frac{\dot a}{a}(\theta_\chi-\theta_b) \nonumber \\ 
=&  \frac{\ddot a}{a}\theta_b -k^2\left(c_b^2\dot \delta_b-\frac{1}{4}\dot \delta_\gamma -\frac{1}{2}\frac{\dot a}{a}\delta_\gamma +\dot \sigma_\gamma + 2\frac{\dot a}{a}\sigma_\gamma\right) \nonumber\\
&-\frac{2\dot a}{a}\dot{\Theta}_{\gamma b}-\frac{2+R}{\tc}\frac{\dot a}{a}\T-\frac{S}{\tb}\frac{\dot a}{a}(\theta_\chi-\theta_b). 
\end{align}

In the first line we used $\frac{\dot a}{a}c_b^2-\dot c_b^2=0$, since in the tight coupling limit $c_b^2\propto T_b \propto a^{-1}$. In the second line, we used Eq.~\ref{ex} to substitute for $\left(\frac{\dot a}{a}\right)^2\theta_b$, and in the third we used Eq.~\ref{photexact} to add and subtract $2\frac{\dot a}{a}\dot \theta_\gamma$.  

Plugging $\dot X$ back into Eq.~\ref{eq1}, we drop the terms involving $\Theta$ and $\sigma_\gamma$, since they are already first order in $\tc$ \cite{Ma:1995}. We get
\begin{align}
\label{eq3}
\dot \Theta_{\gamma \beta}  =& \left(\frac{\dot \tau_c}{\tc}-\frac{2}{1+R}\frac{\dot a}{a}\right)\T  \nonumber \\
&+ \frac{\tc}{1+R}\left[\frac{\ddot a}{a}\theta_b - k^2\left(c_b^2\dot  \delta_b-\frac{1}{4}\dot \delta_\gamma -\frac{1}{2}\frac{\dot a}{a}\delta_\gamma \right) \right. \nonumber\\
& \qquad \left. - \frac{S}{\tb}\left(\frac{\dot a}{a}-\frac{\dot \tau_\chi}{\tb}\right)(\theta_\chi-\theta_b) - \frac{S}{\tb}(\dot \theta_\chi-\dot \theta_b)\right] \nonumber \\
= & \; \Theta_1-\beta\left[(\dot \theta_\chi-\dot \theta_b)+\left(\frac{\dot a}{a}-\frac{\dot \tau_\chi}{\tb}\right)(\theta_\chi-\theta_b)\right],
\end{align}
where $\Theta_1$ is the first-order slip without DM-baryon scattering and $\beta = \frac{S}{1+R}\frac{\tc}{\tb}$.

We see that because of the DM-baryon scattering, the slip derivative contains a remaining factor of $\dot \theta_b$. To get rid of this extra term, we use the exact equation obtained obtained from Eqs.~\ref{ex} and ~\ref{photexact}:
\begin{align}
\label{difftca1}
\dot \theta_b &= -\frac{1}{1+R}\left[\frac{\dot a}{a}\theta_b-c_b^2k^2\delta_b-Rk^2\left(\frac{1}{4}\delta_\gamma-\sigma_\gamma\right)\right. \nonumber\\ 
& \left. -\frac{S}{\tb}(\theta_\chi-\theta_b)+R\dot \Theta_{\gamma \beta} \right]
\end{align}
Plugging the slip derivative Eq.~\ref{eq3} into Eq.~\ref{difftca1}, we collect all the factors of $\dot \theta_b$\ and solve to find the tight-coupling expression for $\dot \theta_b$.
\begin{align}
\label{thetabfinal}
\dot \theta_b =& -\frac{1}{1+R+R\beta}\left[\frac{\dot a}{a}\theta_b-c_b^2k^2\delta_b-Rk^2\left(\frac{1}{4}\delta_\gamma-\sigma_\gamma\right) \right.\nonumber \\
& \quad \left. +R\dot \Theta_1-R\beta\left(\frac{\dot a}{a}-\frac{\dot \tau_\chi}{\tb}\right)(\theta_\chi-\theta_b) \right. \nonumber\\
& \quad \left.-\frac{S}{\tb}(\theta_\chi-\theta_b)-R\beta\dot \theta_\chi\right].
\end{align}

Then, once we have $\dot \theta_b$ in the tight coupling approximation, we use the following exact expression to obtain $\dot \theta_\gamma$.
\begin{align}
\dot \theta_\gamma &= -\frac{1}{R}\left(\dot \theta_b+\frac{\dot a}{a}\theta_b-c_b^2k^2\delta_b^2\right)+k^2\left(\frac{1}{4}\delta_\gamma-\sigma_\gamma\right)\nonumber \\
&+\frac{S}{R\tb}(\theta_\chi-\theta_b).
\end{align}
\end{appendix}

\end{document}